\documentclass[twocolumn,superscriptaddress,amsfont,amssymb,amsmath,showpacs,balancelastpage,nofootinbib,aps,pra,10pt]{revtex4-2}
\usepackage{graphicx,longtable,mathrsfs,color,array}
\usepackage[unicode=true,pdfusetitle,
bookmarks=true,bookmarksnumbered=true,bookmarksopen=true,bookmarksopenlevel=1,
breaklinks=false,pdfborder={0 0 0},backref=false,colorlinks=true]{hyperref}
\hypersetup{citecolor=blue,filecolor=blue,linkcolor=blue,urlcolor=blue,pdfauthor={Name}
}
\usepackage[dvipsnames,table]{xcolor}
\usepackage{amssymb,amsmath,mathtools,mathrsfs,enumitem}
\usepackage{epsfig,subfigure,placeins,float}
\usepackage{booktabs,multirow}
\usepackage{exscale,relsize}
\usepackage[normalem]{ulem}
\usepackage[T1]{fontenc}
\usepackage[utf8]{inputenc}
\usepackage{enumerate}
\usepackage{times, mathptmx}
\usepackage{tikz}
\usepackage{aas_macros}
\usepackage{tabularx}
\usepackage{tabularray}
\usepackage{diagbox}
\usepackage{multirow}
\usepackage{pifont}

\newcolumntype{M}[1]{>{\centering\arraybackslash}m{#1}}
\usepackage{hhline}
\usepackage{makecell}
\usepackage{cleveref}
\usepackage{siunitx}
\usepackage{multirow}
\usepackage{booktabs}
\usepackage{fontawesome5}

\newcolumntype{Y}{>{\centering\arraybackslash}X}

\usetikzlibrary{arrows,positioning,decorations.markings,decorations.pathmorphing,calc}

\newcommand{\be}{\begin{equation}}
\newcommand{\ee}{\end{equation}}
\newcommand{\ba}{\begin{align}}
\newcommand{\ea}{\end{align}}

\newcommand{\comment}[1]{}

\newcolumntype{C}[1]{>{\centering\let\newline\\\arraybackslash\hspace{0pt}}m{#1}}

\newcommand{\nocontentsline}[3]{}
\newcommand{\tocless}[2]{\bgroup\let\addcontentsline=\nocontentsline#1{#2}\egroup}

\def\mpl{M_{\rm Pl}}

\definecolor{hyperref}{RGB}{026,028,087}

\def\gsim{ \lower .75ex \hbox{$\sim$} \llap{\raise .27ex \hbox{$>$}} }
\def\lsim{ \lower .75ex \hbox{$\sim$} \llap{\raise .27ex \hbox{$<$}} }

\newlength{\stheight}
\newcommand\textst[1][fu-grey]{
\ifmmode\setlength{\stheight}{+1.0ex}
\else\setlength{\stheight}{+0.5ex}
\fi
\bgroup\markoverwith{\textcolor{#1}{\rule[\the\stheight]{2pt}{1.0pt}}}\ULon
} 
 
\newcommand{\textins}[2][fu-grey]{
\ifmmode\mathcolor{#1}{#2}
\else\textcolor{#1}{#2}\@\,
\fi
}
\allowdisplaybreaks

\tikzstyle{vecArrow} = [thick, decoration={markings,mark=at position
1 with {\arrow[semithick]{open triangle 60}}},
double distance=1.4pt, shorten >= 5.5pt,
preaction = {decorate},
postaction = {draw,line width=1.4pt, white,shorten >= 4.5pt}]

\begin{document}

\setcounter{tocdepth}{5}
\title{The Symbiosis of Vainshtein and Chameleon Screening}

\author{Sergi Sirera}
 \email{sergi.sirera@port.ac.uk}
 \affiliation{Institute of Cosmology \& Gravitation, University of Portsmouth, Portsmouth, PO1 3FX, U.K.}

\author{Tessa Baker}
 \affiliation{Institute of Cosmology \& Gravitation, University of Portsmouth, Portsmouth, PO1 3FX, U.K.}

 \author{Clare Burrage}
\affiliation{School of Physics and Astronomy, University of Nottingham,
University Park, Nottingham NG7 2RD, UK}

\author{James Hallam}
 \affiliation{Institute of Cosmology \& Gravitation, University of Portsmouth, Portsmouth, PO1 3FX, U.K.}

\author{Krishna Naidoo}
 \affiliation{Institute of Cosmology \& Gravitation, University of Portsmouth, Portsmouth, PO1 3FX, U.K.}
 \affiliation{Max-Planck-Institut f\"ur Astronomie, K\"onigstuhl 17, 69117 Heidelberg, Germany}

\begin{abstract}
We present the first study of coexisting Vainshtein and Chameleon screening in cosmic structures. Screening hides fifth forces locally, keeping modified gravity viable. Realistic dark-energy models can activate several mechanisms at once, yet this interplay has been overlooked. We find that (i) the Chameleon resolves the long-standing Galileon void pathology; (ii) it suppresses or eliminates the Vainshtein radius around massive sources; and (iii) the Vainshtein term shrinks the Chameleon's screened core. These effects reshape the observable signatures central to cosmological tests of gravity.
\end{abstract}

\date{\today}
\maketitle

\section{Introduction} \label{sec:Intro}
While recent cosmological observations seem to provide compelling motivation to modify General Relativity (GR) on large scales \cite{DESI:2024mwx,DES:2025sig},
stringent Solar System tests demand that any viable theory dynamically hide its new degrees of freedom in local dense environments \cite{1993tegp.book.....W,Bertotti:2003rm,Will:2014kxa,Joyce:2014kja,Koyama:2015vza,Burrage:2020jkj,Brax:2021wcv}. This process, known as screening, allows models to transition naturally from cosmological-scale deviations to local GR.
Known screening mechanisms can be broadly divided into two categories: (i) derivative screening (e.g. Vainshtein \cite{Vainshtein:1972sx} and K-mouflage \cite{Babichev:2009ee}), and (ii) environmental screening (e.g. Chameleon \cite{Khoury:2003aq} and Symmetron \cite{Hinterbichler:2010es})\footnote{An additional candidate, Phaedrus screening, has been proposed \cite{Sirera:2026klo}, but remains to be validated.}, each one relying on specific nonlinear operators to suppress scalar interactions.

These mechanisms have traditionally been studied in isolation, but this is an artificial simplification. Generic modified-gravity models can simultaneously activate multiple nonlinear screening operators.
In particular, dark energy models able to accommodate the phantom-divide crossing favoured by recent data feature either or both operator classes~\cite{Ye:2024ywg,Tsujikawa:2025wca,Wolf:2025acj,Linder:2025pqt,Naidoo:2026umv,HiCOLA:James.paper}.
Moreover, combining environmental and derivative operators has been argued to benefit ultraviolet completion~\cite{Nortier:2025gmc}, early-Universe background stability~\cite{Padilla:2015wlv}, and viable dark-energy models~\cite{Ali:2012cv}.\footnote{Combined Vainshtein and K-mouflage screening has been considered in~\cite{Gratia:2016tgq}.}
Spherically symmetric equations containing both operator classes have appeared before~\cite{Kase:2013uja,Noller:2012sv}, but their nonlinear spatial interplay, and its impact on small-scale structure, has largely been ignored.
Ref.~\cite{Sirera:2026klo} took an initial step by deriving a master equation for simultaneous multi-mechanism screening.

Building on this, we study the symbiosis between Vainshtein and Chameleon screening as representatives of the two broad categories.
We introduce an illustrative symbiotic theory, present analytical and numerical solutions for the nonlinear regime, and discuss three phenomenological results: (i) the resolution of the void pathology in pure Galileon models, (ii) the reduction of the Vainshtein screening radius, and (iii) the de-screening (shrinking) of the Chameleon core.
Crucially, we show that this interplay reshapes the spatial fifth-force profiles modelled in $N$-body simulations, fundamentally altering the observable signatures targeted by upcoming Stage IV surveys~\cite{laureijs2011euclid,LSST:2008ijt}.

\section{Mixed screening equation} \label{sec:screening_equation}

Vainshtein and Chameleon are arguably the most well-studied screening mechanisms. The former operates via nonlinear derivative self-interactions, while the latter dynamically increases the scalar's effective mass through (non-derivative) self-interactions, both of which naturally activate in local, high-density environments.
To investigate their interplay, we introduce a minimal Jordan-frame scalar-tensor theory:
\begin{equation}
\begin{split}
S = \int d^4x \sqrt{-g} \bigg[ &\frac{\mpl^2}{2} e^{-2\alpha \phi / \mpl} R - \frac{1}{2}(\partial\phi)^2 \\
& - \frac{\Lambda_2^5 c_2}{\phi} - \frac{c_3}{2\Lambda_3^3} (\partial\phi)^2 \Box\phi \bigg] + S_m,
\end{split}
\label{eq:action}
\end{equation}
where $\phi$ is the scalar field, $S_m$ is the matter action, $R$ is the Ricci scalar, $g$ is the determinant of the metric, $\mpl$ is the reduced Planck mass, and $\Box \equiv \nabla_\mu\nabla^\mu$ is the d'Alembertian operator.
The physical mass scales are matched to the cosmological background as $\Lambda_3^3 \equiv \mpl H_0^2$ and $\Lambda_2^5 \equiv \mpl^3 H_0^2$, where $H_0$ is the value of the Hubble parameter today.
These operators span a vast hierarchy ($H_0$--$\Lambda_3$--$\Lambda_2$--$\mpl$). The classical treatment is nonetheless self-consistent, since the sources lie well below the Galileon strong-coupling scale $\Lambda_3$ (with radius $R\gg\Lambda_3^{-1}\sim10^3{\rm km}$)~\cite{Nicolis:2008in,Deffayet:2009wt}, redressed upward by the Vainshtein mechanism~\cite{Nicolis:2004qq,deRham:2014zqa}.
While the model's radiative stability is not guaranteed (the Chameleon breaks the shift symmetry protecting the Galileon), related shift-symmetry-breaking constructions have been found more radiatively well-behaved than naively expected~\cite{Noller:2018eht,Heisenberg:2020cyi}, and our focus here is the classical phenomenology of the interplay, itself a concrete instance of the broader regulating role additional operators can play~\cite{deRham:2023ngf}.
The dimensionless parameters $\alpha$, $c_2$, and $c_3$ control the strengths of the conformal coupling, Chameleon potential, and Vainshtein kinetic braiding, respectively.
This model thus realises Vainshtein (cubic Galileon term) and Chameleon screening (potential and non-minimal coupling) simultaneously, and is expected to capture the generic phenomenology of their nonlinear interplay regardless of the specific potential or higher-order operators.

To obtain the local spatial dynamics for the scalar, we evaluate the covariant equations of motion on a cosmological Friedmann-Lemaître-Robertson-Walker (FLRW) background. Applying the standard weak-field \cite{Fidler:2017pnb} and quasi-static \cite{Boisseau:2000pr,Esposito-Farese:2000pbo,Copeland:2006wr,Tsujikawa:2007gd} approximations, we express the Ricci scalar in terms of the local matter density and the scalar field's spatial Laplacian (see derivation in Appendix~\ref{app:covariant_eom}). Substituting this back into the scalar equation yields the nonlinear field equation:
\begin{equation}
\begin{split}
    \frac{(1 + 6\alpha^2)}{a^2}&\nabla^2 \phi + \frac{c_3}{a^4 \Lambda_3^3} \left[ (\nabla^2 \phi)^2 - (\nabla_{ij} \phi)^2 \right] \\
    & = \alpha \frac{\rho_m}{\mpl} - \frac{\Lambda_2^5 c_2}{\phi^2},
\label{eq:spatial_master}
\end{split}
\end{equation}
where $\nabla_{ij}\equiv\nabla_i\nabla_j$ and $\nabla^2\equiv\nabla_i\nabla^i$ are spatial derivatives and $a$ is the scale factor.
Here, the linear gradient has absorbed the conformal coupling, shifting the canonical kinetic coefficient; the conformal exponential factor multiplying it is set to unity for the sub-Planckian fields of interest ($\phi\ll\mpl$).
Unlike the perturbative treatment of~\cite{Sirera:2026klo}, which expands the potential about a fixed background, Eq.~\eqref{eq:spatial_master} retains the full scalar, so the field is free to leave its background minimum.

Assuming spherical symmetry, we scale the radial coordinate to the source radius ($x \equiv r/R$) and the scalar field to a fixed reference scale ($y\equiv\phi/\phi_*$), and normalise the equation of motion so that the potential's contribution carries unit coefficient ($-1/y^2$).
Here, $\phi_*$ is a fiducial normalisation (the field value at the effective potential minimum for the reference density $\rho_*=\mpl\Lambda_2^5 c_2/(\alpha\phi_*^2)$).
We model the local density with a top-hat profile, but confirm in Appendix~\ref{app:robustness} the robustness of results to more realistic profiles.
Eq.~\eqref{eq:spatial_master} then reduces to the dimensionless ordinary differential equation:\footnote{We examine linear stability of the spherically symmetric solutions of Eq.~\eqref{eq:dimensionless_spherical} in Appendix~\ref{app:stability}.}
\begin{equation}
    \epsilon \tilde\alpha \frac{(x^2 y')'}{x^2} + 2\beta \frac{[x(y')^2]'}{x^2} = \hat{\rho}(x) - \frac{1}{y^2},
    \label{eq:dimensionless_spherical}
\end{equation}
where primes denote differentiation with respect to $x$.
For a specified density profile $\hat\rho(x)\equiv\rho_m(x)/\rho_*$
(with central and background values denoted $\hat\rho_{\rm in}$ and $\hat\rho_{\rm bg}$), the screening dynamics is governed by three independent dimensionless parameters:
(i) $\epsilon\equiv2(\lambda_*/aR)^2$, the squared ratio of the reference Chameleon Compton wavelength ($\lambda_*^2 = \phi_*^3/(2 \Lambda_2^5 c_2)$, evaluated at the reference amplitude $\phi_*$) to the source radius; the general Compton wavelength
is density-dependent ($\lambda\propto\rho^{-3/4}$), shorter inside a dense source and longer in the dilute background ($\lambda_{\rm in}\ll\lambda_{\rm bg}$);
(ii) the Vainshtein kinetic strength $\beta\equiv c_3 \phi_*^4/(a^4 \Lambda_3^3 \Lambda_2^5 c_2 R^4)$; and
(iii) the conformal coupling $\alpha$, which modifies the linear gradient term via $\tilde\alpha=(1 + 6\alpha^2)$.
Thus $\alpha$ enters solely through $\tilde\alpha$, for which we adopt the representative value $\alpha=1/\sqrt6$ (matched to $f(R)$~\cite{Brax:2008hh}, giving $\tilde\alpha=2$). Different $\tilde\alpha$ choices do not affect the phenomenology, only rescale the thresholds.

\section{Void pathology resolution} \label{sec:void}
\begin{figure}[t]
    \centering
    \includegraphics[width=\linewidth]{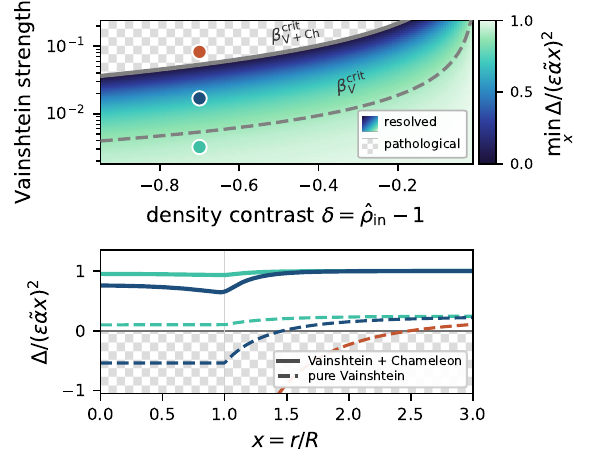}
    \caption{Discriminant $\Delta$ [Eq.~\eqref{eq:Delta}] of the scalar-field gradient [Eq.~\eqref{eq:y_prime}].
    \textbf{Top:} Void pathology ($\Delta<0$) in the plane of density contrast $\delta$ and Vainshtein strength $\beta$ (with $\hat\rho_{\rm bg}=1$, $\epsilon=0.1$).
    Colour shows $\min_x\Delta$ where real.
    The dashed curve is the pure-Vainshtein boundary $\beta^{\rm crit}_{\rm V}$, the solid curve the mixed-screening boundary $\beta^{\rm crit}_{\rm V+Ch}$ ($\min_x\Delta=0$), a factor of $10$ above; between them the Chameleon rescues otherwise-pathological configurations.
    \textbf{Bottom:} radial profile at $\delta=-0.7$ for the three marked configurations: teal, both resolved; blue, Chameleon-rescued; red, both pathological (no real solution).
    }
    \label{fig:void_resolution}
\end{figure}

\begin{figure*}[t]
    \centering
    \includegraphics[width=\linewidth]{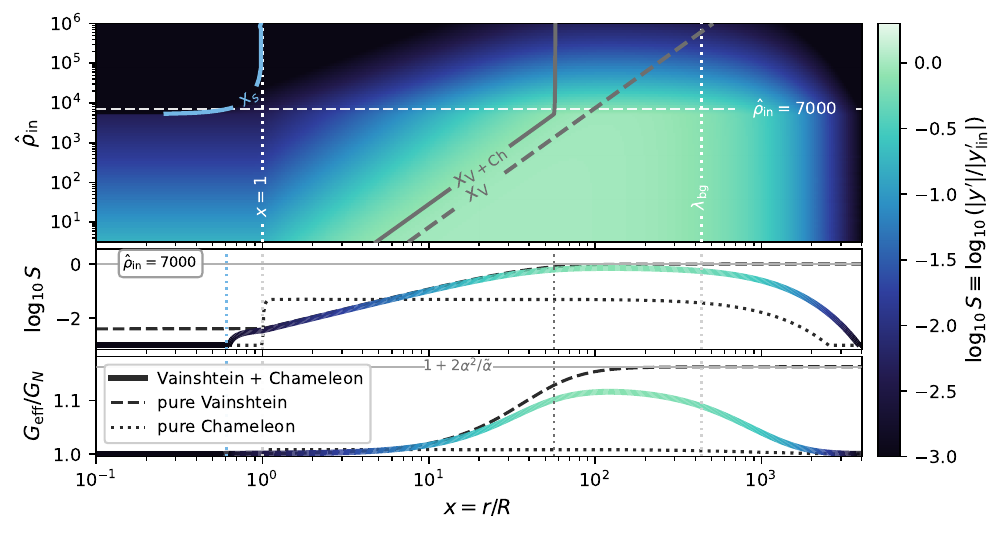}
    \caption{
    \textbf{Top:} Screening factor $S\equiv|y'|/|y'_{\rm lin}|$, from unscreened ($\log_{10}S\simeq0$, pale) to strongly screened ($\log_{10}S<0$, black), against $x$ and $\hat\rho_{\rm in}$; each row is one numerical solution.
    The Chameleon core radius $x_{\rm s}$ (solid blue) appears above a density threshold, pinning the field at its minimum ($y'\approx0$). The symbiotic Vainshtein radius $x_{\rm V+Ch}$ (solid), marking the onset of screening, lies inside its pure-Galileon value $x_{\rm V}$ (dashed). Vertical dotted lines mark the source radius ($x=1$) and the Yukawa cutoff ($\lambda_{\rm bg}$). At fixed $x$, $S$ falls as $\hat\rho_{\rm in}$ grows: the Chameleon-capped charge $\mathcal{I}_{\rm V+Ch}$ saturates (Sec.~\ref{sec:vainshtein_radius}), while the unscreened reference keeps scaling with the full enclosed mass.
    \textbf{Middle:} radial slice at $\hat\rho_{\rm in}=7000$ (white dashed above), crossing all regimes.
    \textbf{Bottom:} the same slice mapped through $G_{\rm eff}/G_N=1+(2\alpha^2/\tilde\alpha)S$, displaying a distinct bump as a signature of this mixed screening.
    The pure-Vainshtein reference differs by panel: $x_{\rm V}$ (top) is the shift-symmetric Galileon; the lower panels retain $\alpha$.
    }
    \label{fig:screening_factor}
\end{figure*}

\begin{figure*}[t]
    \centering
    \includegraphics[width=\textwidth]{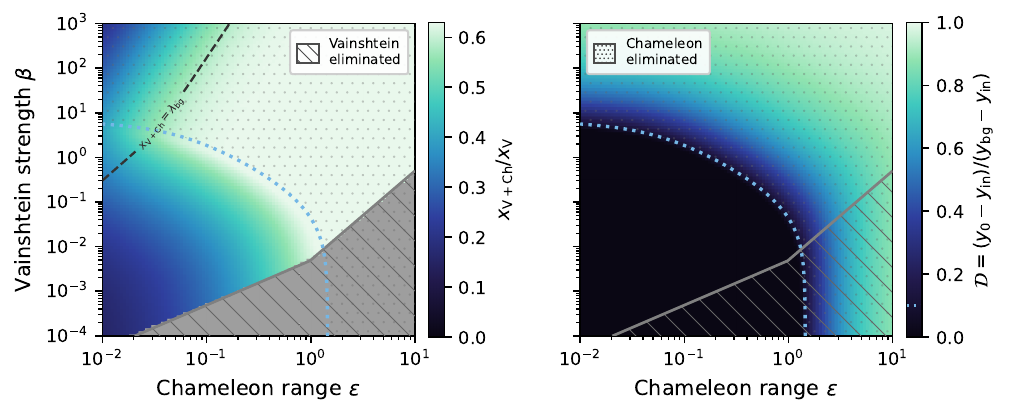}
    \caption{
    Phase diagrams over the plane of Chameleon range $\epsilon=2(\lambda_*/aR)^2$ and Vainshtein strength $\beta$, at $\hat\rho_{\rm in}=300$, $\hat\rho_{\rm bg}=3\times10^{-4}$.
    \textbf{Left:} Vainshtein-radius reduction $x_{\rm V+Ch}/x_{\rm V}$, below unity throughout, saturating at the geometric ceiling $\tilde\alpha^{-2/3}$ (pale) and vanishing below the analytic elimination boundary [Eq.~\eqref{eq:app_two_branch}] ($x_{\rm V+Ch}<1$, grey).
    \textbf{Right:} Chameleon-core de-screening $\mathcal{D}=(y_0-y_{\rm in})/\Delta y$, the central field's displacement from its interior minimum ($\mathcal{D}=0$ screened; $\mathcal{D}=1$ de-screened). At fixed $\epsilon$, larger $\beta$ drives $\mathcal{D}\to1$; the blue dotted curve is $\mathcal{D}=0.1$.
    }
    \label{fig:phase_diagram}
\end{figure*}

Pure Galileon theories, where Vainshtein is the only active screening mechanism, break down critically in underdense environments. While the mechanism merely becomes inefficient in mild cosmic voids \cite{Falck:2014jwa, Barreira:2015vra, Baker:2018mnu}, deep voids trigger an imaginary fifth-force pathology that persists even beyond the quasi-static approximation \cite{Winther:2015pta}.
Within $N$-body simulations, this pathology is bypassed by implementing ad-hoc constant-field cutoffs to prevent numerical breakdown \cite{Winther:2015wla,Wright:2022krq}. Requiring stability in realistic voids therefore seemingly constrains viable parameter space severely \cite{Takadera:2025ehm,Moretti:2026axy,Naidoo:2026umv,HiCOLA:James.paper}.

Here we show that the Chameleon's stable, density-tracking minimum cures the pure-Vainshtein pathology. To see this, we isolate the spatial gradient in Eq.~\eqref{eq:dimensionless_spherical} by integrating from the void centre ($x=0,\ y'(0)=0$) to a radius $x$:
\begin{equation}
    \epsilon \tilde\alpha x^2 y' + 2\beta x (y')^2 = \int_0^x s^2 \left( \hat{\rho} - \frac{1}{y^2} \right) ds \equiv \mathcal{I}(x).
    \label{eq:integrated_master}
\end{equation}
As a quadratic equation for $y'$, taking the positive root for a sensible cosmological limit gives
\begin{equation}
y' = \frac{-\epsilon\tilde\alpha x + \sqrt{\Delta}}{4\beta},
\label{eq:y_prime}
\end{equation}
and the reality of the fifth force then relies strictly on the condition $\Delta\ge0$.
Substituting the enclosed charge [Eq.~\eqref{eq:integrated_master}] back, the discriminant takes the form of a perfect square,
\begin{equation}
    \Delta(x) \equiv (\epsilon\tilde\alpha x)^2 + \frac{8\beta \mathcal{I}(x)}{x} = \big(\epsilon\tilde\alpha x + 4\beta y'\big)^2 .
    \label{eq:Delta}
\end{equation}
This identity is general, so whether a void is pathological is decided by what sources $\mathcal{I}$.
In a pure Vainshtein model, the scalar is sourced by the density perturbation, $\mathcal I_{\rm V}(x)=\int_0^x s^2[\hat\rho(s)-\hat\rho_{\rm bg}]ds$, 
where $\hat\rho_{\rm bg}=\hat\rho(x\to\infty)=\bar\rho_m/\rho_*$ is the far-field value, set by the cosmological mean density $\bar\rho_m$.
Inside a void the core is underdense, $\hat\rho<\hat\rho_{\rm bg}$ (a central density contrast $\delta\equiv\hat\rho_{\rm in}/\hat\rho_{\rm bg}-1<0$), so $\mathcal I_{\rm V}(x)<0$.
For a sufficiently deep void this drives $\Delta<0$ and the gradient $y'$ becomes imaginary.

The Chameleon changes what sources $\mathcal{I}$. It drives the field towards its local density-tracking minimum $y_{\rm v}(x)=\hat\rho^{-1/2}$, defined exactly where the integrand in Eq.~\eqref{eq:integrated_master} vanishes.
By holding the field on this minimum, the Chameleon keeps $y'$ real throughout the void, so $\Delta\ge0$ and the model smoothly bridges the underdensity (Fig.~\ref{fig:void_resolution}). This requires only the existence of a real, density-dependent minimum, not the specific form of the potential, so the resolution is generic: any environmentally-coupled potential admitting such a minimum (e.g. a simple mass term) similarly cures the pathology.

This resolution is not unconditional in $\beta$, however. A sufficiently strong Vainshtein term drives the central field far enough from the interior minimum that the real branch is lost and the pathology re-emerges.
To locate this threshold, we evaluate the discriminant near the void centre, where the density is $\hat\rho_{\rm in}$ and the field is approximately constant (regular at the origin, $y'\to0$), with central value $y_0\equiv y(x\to0)$. Hence, the integrand of Eq.~\eqref{eq:integrated_master} is constant and the enclosed charge is $\mathcal{I}(x)\simeq(\hat\rho_{\rm in}-1/y_0^2)x^3/3$. Inserting this into Eq.~\eqref{eq:Delta} gives $\Delta(x)\simeq x^2\big[(\epsilon\tilde\alpha)^2-\tfrac{8}{3}\beta(1/y_0^2-\hat\rho_{\rm in})\big]$. In a void, the central field lies below the interior minimum ($1/y_0^2>1/y_{\rm in}^2=\hat\rho_{\rm in}$) because the Vainshtein term prevents the field from fully relaxing to it. Raising $\beta$ eventually turns the bracket negative, vanishing at the boundary
\begin{equation}
\beta^{\rm crit}_{\rm V+Ch}\simeq\frac{3(\epsilon\tilde\alpha)^2}{8(1/y_0^2-\hat\rho_{\rm in})}.
\label{eq:void_beta_crit}
\end{equation}
The same estimate for a pure (shift-symmetric) Vainshtein Galileon gives $\beta^{\rm crit}_{\rm V}=3\epsilon^2/[8(\hat\rho_{\rm bg}-\hat\rho_{\rm in})]$.
Relative to this shift-symmetric threshold, the conformal coupling stiffens the kinetic term by a factor $\tilde\alpha^2$, and the Chameleon replaces the background density $\hat\rho_{\rm bg}$ with the smaller central potential term $1/y_0^2$, together raising the threshold by a total factor $\tilde\alpha^2(\hat\rho_{\rm bg}-\hat\rho_{\rm in})/(1/y_0^2-\hat\rho_{\rm in})$. Numerical solutions in Fig.~\ref{fig:void_resolution} confirm this boundary, widening the pathology-free window in $\beta$ by about an order of magnitude, a factor $\tilde\alpha^2$ of which is the normalisation difference.

\section{Reducing the Vainshtein radius} 
\label{sec:vainshtein_radius}

Having established that the Chameleon stabilises underdensities, we now turn to dense sources to see how the two mechanisms interact.
This interplay is displayed in Fig.~\ref{fig:screening_factor}, which maps the numerical screening factor $S\equiv|y'|/|y'_{\rm lin}|$, i.e. the scalar-field gradient normalised to its unscreened linear value (generalising the analytic Vainshtein screening factor of~\cite{Wright:2022krq}).

The Vainshtein radius marks where the nonlinear Galileon term overtakes the linear kinetic one in Eq.~\eqref{eq:Delta},
\begin{equation}
x_{\rm V+Ch}^3 \approx \frac{8\beta\mathcal I_{\rm V+Ch}(x_{\rm V+Ch})}{(\epsilon\tilde\alpha)^2},
\label{eq:xVpCh_def}
\end{equation}
where $x_{\rm V+Ch}$ is the Vainshtein radius of the symbiotic model, different from $x_{\rm V}^3\approx8\beta\hat\rho_{\rm in}/(3\epsilon^2)$ of the pure Galileon~\cite{Nicolis:2008in,Deffayet:2009wt}. The net suppression
\begin{equation}
\Big(\frac{x_{\rm V+Ch}}{x_{\rm V}}\Big)^3\approx\frac{1}{\tilde\alpha^2}\frac{\mathcal I_{\rm V+Ch}}{\mathcal I_{\rm V}}
\label{eq:ratio}
\end{equation}
separates cleanly into a fixed ceiling $\tilde\alpha^{-2/3}\leq1$ and a dynamical charge reduction $(\mathcal I_{\rm V+Ch}/\mathcal I_{\rm V})^{1/3}\le1$.
The Chameleon freezes the screened interior at the minimum $y_{\rm in}=\hat\rho_{\rm in}^{-1/2}$, which sources no gradient, so the field only has to climb the excursion $\Delta y\equiv y_{\rm bg}-y_{\rm in}$ to reach the background. Two charges then compete to set the Vainshtein radius: the excursion charge $\mathcal I_{\rm exc}=\epsilon\tilde\alpha\Delta y$ this climb demands (Appendix~\ref{app:vainshtein}), and the enclosed mass $\mathcal I_{\rm V}=(\hat\rho_{\rm in}-\hat\rho_{\rm bg})/3\simeq\hat\rho_{\rm in}/3$
physically available. Since the field sources only the charge needed to reach the background, but cannot source more than its contained mass, the effective charge is the smaller of the two, $\mathcal I_{\rm V+Ch}=\min(\mathcal I_{\rm exc},\mathcal I_{\rm V})$. When the background is dilute the excursion is the binding constraint ($\mathcal I_{\rm exc}<\mathcal I_{\rm V}$), and the charge, hence the sphere, drops below its pure-Galileon size. The outcome falls into three regimes (Appendix~\ref{app:vainshtein}), shown in the left panel of Fig.~\ref{fig:phase_diagram}: (i) a reduced but still macroscopic sphere; (ii) a saturation ceiling set by $\tilde\alpha$; and (iii) complete elimination.

In (i), the still macroscopic sphere ($x_{\rm V+Ch}>1$) shrinks below its pure-Galileon size to
\begin{equation}
x_{\rm V+Ch}\simeq\left(\frac{4\beta\Delta y}{G_0\epsilon\tilde\alpha}\right)^{1/2},\qquad G_0=\frac{\Gamma(1/6)\Gamma(1/3)}{4\sqrt{\pi}}\simeq2.10.
\label{eq:xVpCh_red}
\end{equation}
In (ii), once the excursion charge reaches the full mass ($\mathcal I_{\rm exc}\ge\mathcal I_{\rm V}$), the charge reduction switches off and the ratio Eq.~\eqref{eq:ratio} is saturated at the ceiling $\tilde\alpha^{-2/3}$. In (iii), when the coupling is too weak to push the sphere past the source it is drawn inside it ($x_{\rm V+Ch}<1$) and Vainshtein screening disappears entirely.
Setting $x_{\rm V+Ch}=1$ gives a two-branch boundary [Eq.~\eqref{eq:app_two_branch}] (confirmed numerically in Fig.~\ref{fig:phase_diagram}): an excursion-limited $8\beta\Delta y\lesssim\epsilon\tilde\alpha$ with slope $\propto\epsilon$, crossing over to a steeper mass-capped branch $\propto\epsilon^2$ at higher $\epsilon$.
Since $\epsilon\propto R^{-2}\rho_*^{-3/2}$, at fixed $\rho_*$ the $\epsilon$ axis tracks source size; changing $\rho_*$ rescales the $(\epsilon,\beta)$ axes, not the physics.

\section{De-screening the Chameleon core}
\label{sec:chameleon}

Having seen how the Chameleon reshapes the Vainshtein sector, we now examine the reciprocal effect. In a pure Chameleon model, a screened body has a sharp screening transition: the field is pinned at the high-density minimum $y_{\rm in}$ throughout the interior (where the Compton wavelength $\lambda_{\rm in} \ll R$), and climbs to the background $y_{\rm bg}$ within a few source radii outside it, following a $1/x$ Coulomb tail. We show that including the Vainshtein term de-screens (shrinks) that core, which we quantify through the de-screening function
\begin{equation}
    \mathcal{D}\equiv\frac{y_0 - y_{\rm in}}{y_{\rm bg} - y_{\rm in}},
    \label{eq:Ddef}
\end{equation}
where $\mathcal{D}\simeq0$ for a fully screened core sitting at its interior minimum, and $\mathcal{D}>0$ once the central field is lifted from the potential minimum, i.e. the Chameleon fails to activate.

Within a macroscopic Vainshtein sphere, the nonlinear kinetic term caps the spatial gradient [Eq.~\eqref{eq:y_prime}], flattening the field across $x\lesssim x_{\rm V+Ch}$, so the interior can no longer reach $y_{\rm in}$ and $\mathcal{D}$ grows. The right panel of Fig.~\ref{fig:phase_diagram} maps $\mathcal{D}$ across the $(\epsilon,\beta)$ plane, showing that at fixed $\epsilon$, increasing $\beta$ drives the core from screened ($\mathcal{D}\simeq0$) to eliminated ($\mathcal{D}\to1$).

To locate the onset of this elimination, we assume a screened core solution exists and match it to the transition region and the exterior Coulomb tail. This gives a screened-core radius $x_{\rm s}^2=1-2\epsilon\tilde\alpha\Delta y/\hat\rho_{\rm in}$, which shrinks to zero at (Appendix~\ref{app:core-threshold})
\begin{equation}
\epsilon_{\rm crit}\simeq\frac{\hat\rho_{\rm in}\hat\rho_{\rm bg}^{1/2}}{2\tilde\alpha}\approx1.3,
\label{eq:eps_crit}
\end{equation}
with the numerical value evaluated at the fixed densities of Fig.~\ref{fig:phase_diagram}. The Vainshtein term relaxes this bound: a larger $\beta$ de-screens the core at smaller $\epsilon$, so the elimination boundary (blue dotted, $\mathcal{D}=0.1$) bends leftwards in Fig.~\ref{fig:phase_diagram}.
Where the core remains screened, the same gradient cap also widens the screening transition, stretching its radial width towards the Vainshtein scale $x_{\rm V+Ch}$.

\section{Conclusions}
\label{sec:Conclusions}
Screening mechanisms, essential for the viability of dark energy models, have historically been treated in isolation. This represents an unnatural simplification in general scalar-tensor theories capable of realising multiple mechanisms simultaneously, as is true for many recently proposed dynamical dark energy models.
Here, we have derived and solved a spatial master equation to analyse the nonlinear symbiosis between Vainshtein and Chameleon screening, showing that this interplay fundamentally alters the scalar field's kinematics.
Specifically, the Chameleon's density-tracking minimum resolves the imaginary fifth-force pathology inherent to pure Galileon models, restoring stability deep within cosmic voids.
In dense environments, the Chameleon shields the core of massive sources, pulling the Vainshtein radius inward and, eventually, eliminating it; conversely, the Vainshtein operator's gradient cap de-screens the Chameleon core, broadening its screening transition where it survives.

These reciprocal effects carry profound implications for next generation cosmological probes. Precision tests of gravity with Stage-IV surveys rely on accurate $N$-body models, and assuming isolated screening mechanisms will systematically mischaracterise the fifth-force profile in theories that realise multiple screenings.
These results can similarly impact laboratory bounds. Since the Vainshtein term lifts the Chameleon core, exclusion regions mapped for isolated Chameleons~\cite{Burrage:2017qrf} need not transfer to theories realising both mechanisms.
Ultimately, by rescuing Galileon stability in voids and altering structural signatures around dark matter halos, this symbiosis reopens previously discarded regions of parameter space as a theoretically robust, observationally distinct target for future surveys.

\begin{acknowledgments}
We thank Philippe Brax, Claudia de Rham, Kazuya Koyama and Johannes Noller for useful discussions.
SS, TB and KN are supported by ERC Starting Grant SHADE (grant no. StG 949572). TB is further supported
by a Royal Society University Research Fellowship (grant no.~URF\textbackslash R\textbackslash 231006). CB is supported in part by STFC Consolidated Grants (grant nos. ST/T000732/1 and ST/C004235/1). JH is supported by a PhD studentship from UKRI-STFC.
We have used \href{https://github.com/sergisl/escut}{\faGithub \;escut}~\cite{escut} and \href{https://github.com/sergisl/xAlpha}{\faGithub \;xAlpha}~\cite{xAlpha}, where we include all the files to reproduce the results of this paper.
The authors are grateful for computational resources on the Sciama High Performance Computing (HPC) cluster, which is supported by the Institute of Cosmology and Gravitation, the South-East Physics Network (SEPNet) and the University of Portsmouth.
For the purpose of open access, the authors have applied a Creative Commons Attribution (CC BY) licence to any Author Accepted Manuscript version arising from this work.
\end{acknowledgments}

\appendix

\section{Covariant equations and the spatial master equation}
\label{app:covariant_eom}
We take the action of Eq.~\eqref{eq:action}, used throughout the main text, to be a minimal laboratory for the screening interplay rather than a complete dark-energy model. We retain the canonical kinetic sign since the Chameleon potential removes the need for the ``wrong'' sign pure Galileons often require to accelerate~\cite{DeFelice:2011bh,Barreira:2013jma,Melville:2022ykg}. While the cubic Galileon is ruled out as a self-accelerating dark-energy model~\cite{Renk:2017rzu,Kable:2021yws,Seraille:2024beb}, it underpins viable constructions~\cite{Atayde:2024tnr,Wolf:2025acj,Naidoo:2026umv,HiCOLA:James.paper} and is to date the only theory admitting stable black holes with cosmological hair, testable by ringdown~\cite{Smulders:2026qwc,Smulders:2026bya}.

Varying the action in Eq.~\eqref{eq:action} with respect to the metric $g_{\mu\nu}$ yields the covariant Einstein equation. Taking the trace of this equation and assuming a non-relativistic matter source ($T^{(m)} \approx -\rho_m$) gives
\begin{equation}
    -\mpl e^{-2\alpha\phi/\mpl}\left(\mpl R + 6\alpha\Box\phi\right) + \mathcal{O}\left((\nabla\phi)^2\right) = -\rho_m.
\end{equation}
In the sub-horizon quasi-static limit, the scalar's own stress-energy trace is negligible. Being sub-Planckian ($\phi\ll\mpl$), its $\mathcal{O}((\nabla\phi)^2)$ contributions are dimensionally suppressed relative to the $\sim\mpl\Box\phi$ term.
Approximating the d'Alembertian by its spatial gradients ($\Box \approx \nabla^2/a^2$), the Ricci scalar evaluates to
\begin{equation}
    R \approx \frac{\rho_m}{\mpl^2}e^{2\alpha\phi/\mpl} - \frac{6\alpha}{\mpl a^2}\nabla^2\phi.
    \label{eq:ricci_trace}
\end{equation}

Varying the action with respect to $\phi$ yields the covariant scalar equation of motion
\begin{equation}
\begin{split}
    -\alpha \mpl e^{-2\alpha\phi/\mpl} R + \frac{\Lambda_2^5 c_2}{\phi^2} + \Box\phi \\
    + \frac{c_3}{\Lambda_3^3} \left[ (\Box\phi)^2 - (\nabla_\mu\nabla_\nu\phi)^2 \right] = 0,
\end{split}
    \label{eq:covariant_scalar}
\end{equation}
where, in the sub-horizon quasi-static limit, we have dropped the curvature term $R_{\mu\nu}\nabla^\mu\phi,\nabla^\nu\phi$ arising in the covariant cubic-Galileon field equation, suppressed relative to the retained terms.
Substituting Eq.~\eqref{eq:ricci_trace} into Eq.~\eqref{eq:covariant_scalar}, the conformal exponential factor cancels against the matter source and multiplies the $6\alpha^2$ gradient term; setting it to unity for sub-Planckian fields, the linear spatial gradient shifts by $(1 + 6\alpha^2)$ and we recover Eq.~\eqref{eq:spatial_master}.

\section{Perturbative stability}
\label{app:stability}

\begin{figure}[t]
\centering
\includegraphics[width=\linewidth]{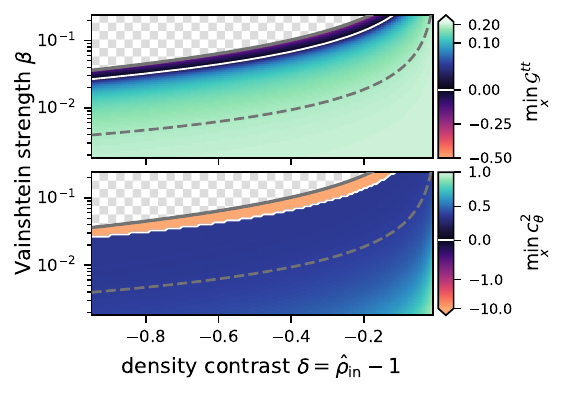}
\caption{Perturbative stability on the void $(\delta,\beta)$ plane, same grid and boundary curves ($\beta^{\rm crit}_{\rm V}$ dashed, $\beta^{\rm crit}_{\rm V+Ch}$ solid) as Fig.~\ref{fig:void_resolution}.
\textbf{Top:} no-ghost condition, Eq.~\eqref{eq:app_noghost}.
\textbf{Bottom:} angular speed condition, Eq.~\eqref{eq:app_cth}. The chequered region ($\Delta<0$) has no real solution and is excluded.
The white contour marks the ghost/gradient boundary ($\mathcal{G}^{tt}=0$,$c_\theta^2=0$), showing that $\Delta$-reality alone does not guarantee stability.}
\label{fig:void_stability}
\end{figure}

\begin{figure*}[t]
\centering
\includegraphics[width=\textwidth]{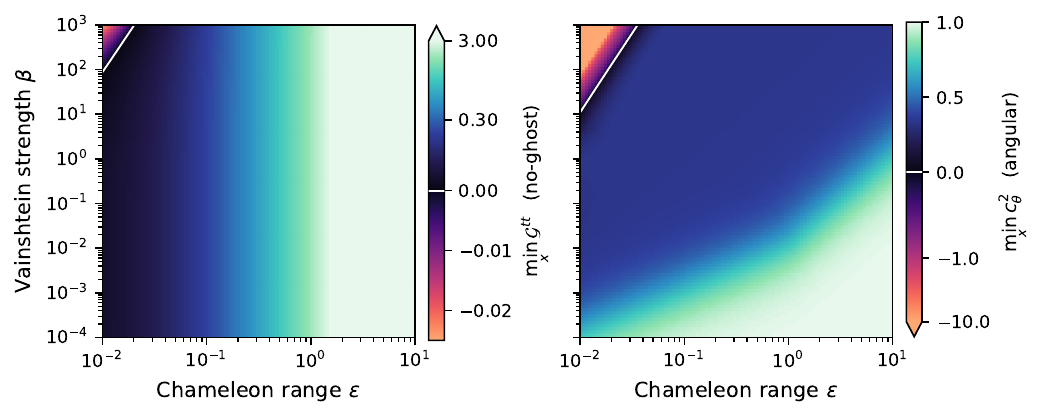}
\caption{Perturbative stability over the same $(\epsilon,\beta)$ plane as Fig.~\ref{fig:phase_diagram}, at $\hat\rho_{\rm in}=300$, $\hat\rho_{\rm bg}=3\times10^{-4}$ and $\tilde\alpha=2$; each pixel is one numerical solution. \textbf{Left:} the no-ghost function $\min_x\mathcal G^{tt}$, Eq.~\eqref{eq:app_noghost}.
\textbf{Right:} the minimum angular sound speed $\min_x c_\theta^2$, Eq.~\eqref{eq:app_cth}. Cold colours ($\ge0$) are stable, warm colours ($<0$) unstable; in the linear regime $c_\theta^2\to1$ and $\mathcal G^{tt}\to\epsilon\tilde\alpha$. The white line marks the stability boundary ($\mathcal G^{tt}=0$, $c_\theta^2=0$). The solutions are ghost-free and gradient-stable everywhere except the thin top-left wedge (the pure cubic-Galileon limit) in which the Chameleon is effectively absent.}
\label{fig:stability}
\end{figure*}

We analyse linear fluctuations about the static, spherically symmetric solutions of Eq.~\eqref{eq:dimensionless_spherical}. In the weak-field limit it suffices to perturb the scalar, $y\to y+\pi$, in the effective Lagrangian whose equation of motion is Eq.~\eqref{eq:dimensionless_spherical},
\begin{equation}
\mathcal L_{\rm eff}=-\tfrac12\epsilon\tilde\alpha(\partial y)^2-\tfrac{\beta}{2}(\partial y)^2\Box y-V(y).
\end{equation}
At second order, $\mathcal L^{(2)}=-\tfrac12\epsilon\tilde\alpha(\partial\pi)^2-\tfrac{\beta}{2}(\partial\pi)^2\Box y-\beta(\partial y\!\cdot\!\partial\pi)\Box\pi$. Integrating the last term by parts, the fluctuation propagates on the effective inverse metric
\begin{equation}
\mathcal G^{\mu\nu}=-\epsilon\tilde\alpha g^{\mu\nu}+2\beta\big(\nabla^\mu\nabla^\nu y-\Box yg^{\mu\nu}\big),
\label{eq:app_Gmetric}
\end{equation}
the cubic Galileon supplying the background-dependent term. For the static background ($\Box y=(x^2y')'/x^2$) its components give the no-ghost condition and the radial and angular sound speeds,
\begin{align}
\mathcal G^{tt}&=\epsilon\tilde\alpha+2\beta\Box y,\label{eq:app_noghost}\\
c_r^2&=-\frac{\mathcal G^{rr}}{\mathcal G^{tt}}=\frac{\epsilon\tilde\alpha x+4\beta y'}{x\mathcal G^{tt}}=\frac{\sqrt\Delta}{x\mathcal G^{tt}},\label{eq:app_cr}\\
c_\theta^2&=\frac{\epsilon\tilde\alpha+2\beta(y''+y'/x)}{\mathcal G^{tt}}.\label{eq:app_cth}
\end{align}
The solutions are stable where all three conditions hold: $\mathcal G^{tt}>0$, $c_r^2\geq0$,  and $c_\theta^2\geq0$. Radial stability is governed by the same discriminant $\Delta$ of Eq.~\eqref{eq:Delta}, so resolving the void pathology ($\Delta\ge0$) is necessary to restore radial stability, in addition to requiring $\mathcal G^{tt}>0$. Evaluating Eqs.~\eqref{eq:app_noghost}--\eqref{eq:app_cth} on the numerical solutions across the Fig.~\ref{fig:stability} parameter space, all three conditions are satisfied wherever the Chameleon is active. Only in the deep-Galileon corner (small $\epsilon$, large $\beta$), where the model reduces to the pure Galileon, do the ghost and the angular instability re-emerge, recovering the known pure-Galileon pathologies~\cite{Andrews:2010km,Koyama:2013paa}. On the void plane (Fig.~\ref{fig:void_stability}), a narrow band of configurations near $\beta^{\rm crit}_{\rm V+Ch}$ remains $\Delta$-real yet ghost-unstable ($\mathcal G^{tt}<0$).
Both $c_r^2$ and $c_\theta^2$ reduce to unity in the linear regime, as required.

\section{Reduction and elimination of the Vainshtein radius}
\label{app:vainshtein}

We derive here the results quoted in Sec.~\ref{sec:vainshtein_radius}. For a deeply-screened top-hat halo, the Chameleon pins the field at $y_{\rm in}$ throughout the interior, so the entire excursion to the background occurs outside the source. Integrating the exact gradient Eq.~\eqref{eq:y_prime} from the surface outward fixes the excursion $\Delta y\equiv y_{\rm bg}-y_{\rm in}$ to\footnote{Integrating from the surface gives $y_{\rm bg}-y(1)$; the field departs from $y_{\rm in}$ by only $\mathcal{O}(\epsilon)$ across the surface layer, so $y(1)\simeq y_{\rm in}$.}
\begin{equation}
\Delta y=\frac{\epsilon\tilde\alpha x_{\rm V+Ch}^2}{4\beta}G\!\left(\frac{1}{x_{\rm V+Ch}}\right),
\label{eq:app_excursion}
\end{equation}
where we have used
\begin{equation}
G(a)\equiv\int_a^\infty\!\Big[\sqrt{u^2+\tfrac1u}-u\Big]du,
\label{eq:app_G}
\end{equation}
and $u\equiv x/x_{\rm V+Ch}$ is the radius in units of the Vainshtein radius. The integrand is the Vainshtein enhancement of the gradient over the linear one, i.e. significant only inside the sphere ($u<1$) and vanishing outside. The lower limit $a=1/x_{\rm V+Ch}=R/r_{\rm V+Ch}$ measures how deeply the source sits inside its own Vainshtein sphere, and its two limits correspond to different phenomenology.

\emph{Reduction} ($a\to0$, $x_{\rm V+Ch}\gg1$). The source sits deep inside a large Vainshtein sphere, so the full nonlinear region out to $x_{\rm V+Ch}$ contributes to the excursion and $G\to G(0)=\Gamma(1/6)\Gamma(1/3)/(4\sqrt\pi)\equiv G_0\simeq2.10$, with $\Gamma$ the gamma function. Solving Eq.~\eqref{eq:app_excursion} then gives Eq.~\eqref{eq:xVpCh_red}, showing that the radius grows as $\beta^{1/2}$ yet remains inside the pure-Galileon radius ($\propto\beta^{1/3}$) throughout, saturating at the geometric ceiling $\tilde\alpha^{-2/3}$ (a fixed offset from the shift-symmetric comparison theory's normalisation).

\emph{Elimination} ($a\to1$, $x_{\rm V+Ch}\to1$). The Vainshtein sphere has contracted to the source surface, so the exterior is essentially linear and $G\to1/(2a)$, its linear-tail value (already accurate at $a=1$, where $G(1)\simeq1/2$). Equation~\eqref{eq:app_excursion} then reduces to $x_{\rm V+Ch}^3\approx8\beta\Delta y/(\epsilon\tilde\alpha)$, so the mechanism is eliminated once $8\beta\Delta y\lesssim\epsilon\tilde\alpha$ holds.

Finally, the enclosed charge cannot exceed the full mass, so $\mathcal I_{\rm V+Ch}=\min(\mathcal I_{\rm exc},\mathcal I_{\rm V})$, with $\mathcal I_{\rm exc}=\epsilon\tilde\alpha\Delta y$ the excursion charge and $\mathcal I_{\rm V}\simeq\hat\rho_{\rm in}/3$ the mass. Setting $x_{\rm V+Ch}=1$ in Eq.~\eqref{eq:xVpCh_def} therefore yields a two-branch elimination boundary,
\begin{equation}
\beta_{\rm elim}=\max\!\left(\frac{\epsilon\tilde\alpha}{8\Delta y},\ \frac{3(\epsilon\tilde\alpha)^2}{8\hat\rho_{\rm in}}\right),
\label{eq:app_two_branch}
\end{equation}
whose excursion-limited slope $\propto\epsilon$ and mass-capped slope $\propto\epsilon^2$ cross where $\epsilon\tilde\alpha\Delta y=\hat\rho_{\rm in}/3$. This is the grey boundary in the left panel of Fig.~\ref{fig:phase_diagram}.
Note that $\mathcal I_{\rm exc}=\epsilon\tilde\alpha\Delta y$ is the $a\to1$ (linear-tail) limit of Eq.~\eqref{eq:app_excursion}, appropriate for locating $x_{\rm V+Ch}=1$; the deep-Vainshtein regime is instead governed by the $a\to0$ limit, Eq.~\eqref{eq:xVpCh_red}, whose $\beta^{1/2}$ growth is cut off once the mass cap saturates the ratio at $\tilde\alpha^{-2/3}$. The map itself is free of either assumption: $x_{\rm V+Ch}$ is measured from each numerical solution, and only the overlaid analytic boundary assumes a pinned core.

\emph{The pure-Vainshtein comparison.} We quote $x_{\rm V}$ for a shift-symmetric cubic Galileon, sourced in the standard way by a direct linear coupling to the matter trace of the same strength $\alpha$. Its field equation is Eq.~\eqref{eq:dimensionless_spherical} with $\tilde\alpha\to1$ and the uncapped charge $\mathcal I_{\rm V}\simeq\hat\rho_{\rm in}/3$. We keep the conformal coupling out of this baseline because it is constitutive of the Chameleon rather than of Vainshtein. Without it, the effective potential has no density dependence, and no Chameleon exists. This is a convention, and it is not unique. Setting only $c_2=0$ in Eq.~\eqref{eq:action} instead leaves a cubic Galileon which, canonically normalised, carries $\alpha_{\rm eff}=\alpha/\sqrt{\tilde\alpha}$ and $\Lambda_3^3\to\tilde\alpha^{3/2}\Lambda_3^3$, its Vainshtein radius is Eq.~\eqref{eq:xVpCh_def} with $\tilde\alpha$ retained, and the ratio Eq.~\eqref{eq:ratio} then loses its $\tilde\alpha^{-2/3}$ ceiling altogether. The $\tilde\alpha$ factors in Eqs.~\eqref{eq:void_beta_crit} and~\eqref{eq:ratio} are therefore normalisation differences between comparison theories rather than dynamical effects of the Chameleon potential. The genuinely dynamical effect is the charge reduction $\mathcal I_{\rm V+Ch}/\mathcal I_{\rm V}$. For the same reason, the lower panels of Fig.~\ref{fig:screening_factor} retain $\alpha$, as there $G_{\rm eff}/G_N$ is a prediction of the theory itself, fixed by $\alpha_{\rm eff}^2=\alpha^2/\tilde\alpha$.

\section{The Chameleon core de-screening threshold}
\label{app:core-threshold}

We derive the pure-Chameleon elimination threshold quoted in Eq.~\eqref{eq:eps_crit}. Setting $\beta=0$, the master equation reduces to $\epsilon\tilde\alpha(x^2y')'/x^2=\hat\rho(x)-1/y^2$, with a top-hat source ($\hat\rho=\hat\rho_{\rm in}$ for $x<1$, $\hat\rho_{\rm bg}$ for $x>1$) and minima $y_{\rm in}=\hat\rho_{\rm in}^{-1/2}$, $y_{\rm bg}=\hat\rho_{\rm bg}^{-1/2}$. A screened body admits a thin-shell solution in three regions: a pinned core ($y=y_{\rm in}$, $y'=0$) for $x<x_{\rm s}$; a rolling shell for $x_{\rm s}<x<1$, where the potential term is negligible against the matter term so that $\epsilon\tilde\alpha(x^2y')'/x^2\simeq\hat\rho_{\rm in}$ with general solution $y=\hat\rho_{\rm in}x^2/(6\epsilon\tilde\alpha)+a/x+b$; and an exterior Coulomb tail $y=y_{\rm bg}+c/x$ for $x>1$ (valid since $\lambda_{\rm bg}\gg1$).

Regularity at the core edge, $y'(x_{\rm s})=0$ and $y(x_{\rm s})=y_{\rm in}$, fixes $a=\hat\rho_{\rm in}x_{\rm s}^3/(3\epsilon\tilde\alpha)$ and $b=y_{\rm in}-\hat\rho_{\rm in}x_{\rm s}^2/(2\epsilon\tilde\alpha)$; continuity of $y'$ at the surface fixes $c=\hat\rho_{\rm in}(x_{\rm s}^3-1)/(3\epsilon\tilde\alpha)$. Continuity of $y$ at $x=1$ then gives
\begin{equation}
\Delta y\equiv y_{\rm bg}-y_{\rm in}=\frac{\hat\rho_{\rm in}}{2\epsilon\tilde\alpha}\big(1-x_{\rm s}^2\big),
\label{eq:app_shell}
\end{equation}
i.e. $x_{\rm s}^2=1-2\epsilon\tilde\alpha\Delta y/\hat\rho_{\rm in}$. The screened core shrinks as $\epsilon$ grows and disappears ($x_{\rm s}\to0$) at
\begin{equation}
\epsilon_{\rm crit}=\frac{\hat\rho_{\rm in}}{2\tilde\alpha\Delta y}\simeq\frac{\hat\rho_{\rm in}\hat\rho_{\rm bg}^{1/2}}{2\tilde\alpha},
\label{eq:app_eps_crit}
\end{equation}
using $\Delta y\simeq y_{\rm bg}=\hat\rho_{\rm bg}^{-1/2}$. For $\hat\rho_{\rm in}=300$, $\hat\rho_{\rm bg}=3\times10^{-4}$, and $\tilde\alpha=2$, this gives $\epsilon_{\rm crit}\approx1.3$. Restoring a finite $\beta$ flattens the interior gradient and lifts the core, lowering $\epsilon_{\rm crit}$, as traced by the boundary in Fig.~\ref{fig:phase_diagram}.

\section{Numerical method}

\begin{figure}[t]
\centering
\includegraphics[width=\linewidth]{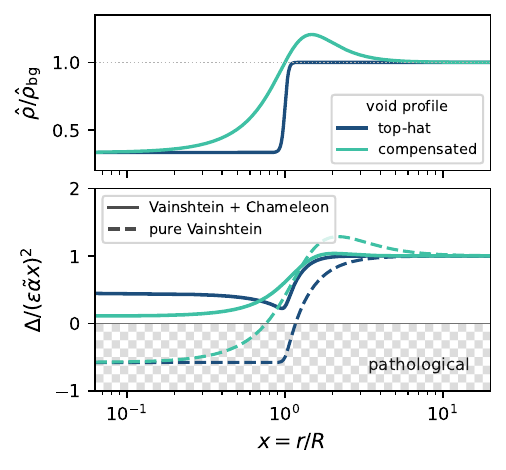}
\caption{Robustness of the void resolution. \textbf{Top:} density profiles for an idealised top-hat and a compensated void with an overdense ridge. \textbf{Bottom:} radial discriminant profile for the symbiotic solution (solid), which stays $\geq0$ for both profiles; and the pure Vainshtein (dashed) solution, which is pathological ($\Delta<0$) in the core. The resolution is generic to the profile shape.}
\label{fig:robustness_void}
\end{figure}

\begin{figure*}[t]
\centering
\includegraphics[width=\textwidth]{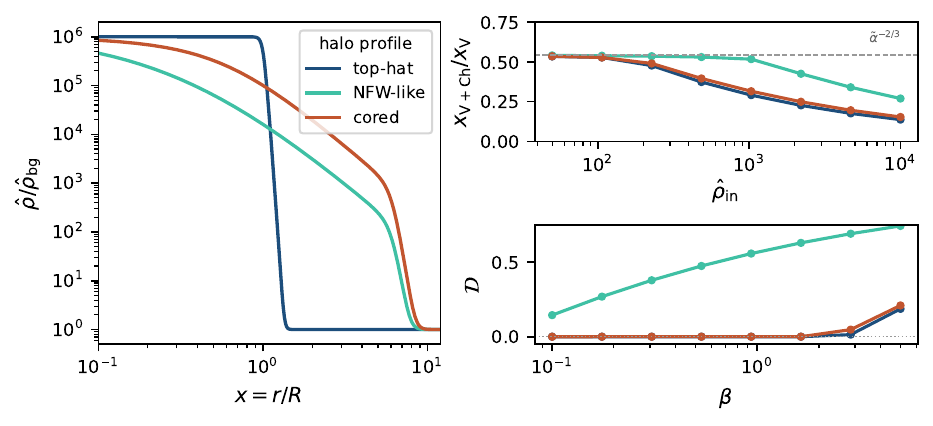}
\caption{Robustness of the halo results for three profiles. \textbf{Left:} the density profiles: top-hat, NFW-like (a regularised cored-NFW, finite at the centre), and cored (Burkert). \textbf{Top right:} Vainshtein-radius reduction $x_{\rm V+Ch}/x_{\rm V}$ versus central density $\hat\rho_{\rm in}$ at fixed $\beta$. We see it remains below unity throughout, equal to the geometric value $\tilde\alpha^{-2/3}$ at low density and deepening with density. \textbf{Bottom right:} Chameleon-core de-screening $\mathcal{D}$ [Eq.~\eqref{eq:Ddef}] versus Vainshtein strength $\beta$ at fixed $\hat\rho_{\rm in}$. The effect is generic to all profiles, with the flat-core profiles retaining a threshold and the NFW-like cusp de-screening continuously.}
\label{fig:robustness_halo}
\end{figure*}

We solve the dimensionless master equation [Eq.~\eqref{eq:dimensionless_spherical}] using the \href{https://github.com/sergisl/escut}{\faGithub \;escut}~\cite{escut} solver. We compute this as a first-order boundary-value problem in the pair $(y,\mathcal I)$, where $\mathcal I(x)$ is the enclosed charge [Eq.~\eqref{eq:integrated_master}]. Differentiating yields
\begin{align}
\frac{d\mathcal I}{dx}&=x^2\!\left(\hat\rho(x)-\frac{1}{y^2}\right),\\
\frac{dy}{dx}&=\frac{-\epsilon\tilde\alpha x+\sqrt{(\epsilon\tilde\alpha x)^2+8\beta\mathcal I/x}}{4\beta},
\end{align}
the second relation being Eq.~\eqref{eq:y_prime}. This form avoids the singular denominator of the second-order equation for $y$ and keeps the smooth, physical charge $\mathcal I$ as a dependent variable.
Throughout, we set the conformal exponential factor to unity, which reproduces $y'(x)$ to better than $0.5\%$ for $2\alpha\phi_{\rm bg}/\mpl\lesssim0.01$ and to $\sim6\%$ even at $2\alpha\phi_{\rm bg}/\mpl=0.1$.
Regularity at the centre imposes $\mathcal I(x_{\min})=0$ (equivalently $y'(x_{\min})=0$); at the outer edge we impose a Robin (Yukawa-decay) condition, $y'(x_{\max})+(m_{\rm bg}+1/x_{\max})(y-y_{\rm bg})=0$, with 
$m_{\rm bg}^2=2/(\epsilon\tilde\alpha y_{\rm bg}^3)$ the background Chameleon mass (linearising the potential about its density-tracking minimum $y_{\rm v}=\hat\rho^{-1/2}$ gives $m_{\rm v}^2=2\hat\rho^{3/2}/(\epsilon\tilde\alpha)$, so $\lambda_{\rm v}=m_{\rm v}^{-1}\propto\hat\rho^{-3/4}$) so the field relaxes to the background.

We integrate the system with the collocation solver \texttt{solve\_bvp}~\cite{2020SciPy-NMeth}, seeded by a minimum-tracking initial guess $y(x)=y_{\rm v}(\hat\rho(x))$ with the consistent $\mathcal I$, and continued in the Vainshtein strength: $\beta$ is ramped from $0$ (the readily solved pure-Chameleon limit) to its target over several homotopy steps, each seeded by the previous solution. Deeply screened cores and deep voids, where the field spans several decades, are solved in the logarithmic variable $u=\ln y$.

\section{Robustness to the density profile}
\label{app:robustness}
The main-text results use a smoothed top-hat source. Here, we confirm that the three effects persist for realistic density profiles, matched to the same central and background densities.

\emph{Void resolution} (Fig.~\ref{fig:robustness_void}). We compare the top-hat with a compensated void~\cite{Hamaus:2014fma}, whose overdense ridge at the edge renders the enclosed charge non-monotonic. For both profiles the pure-Vainshtein discriminant turns negative in the core (an imaginary fifth force), whereas the Vainshtein+Chameleon solution keeps $\Delta\ge0$ throughout. The resolution is therefore generic: it requires only a real density-tracking minimum, not the top-hat shape.

For halos we test two realistic shapes alongside the top-hat: an NFW-like profile~\cite{Navarro:1996gj}, $\rho\propto[(r/r_s)(1+r/r_s)^2]^{-1}$, the cuspy form predicted by cold-dark-matter $N$-body simulations (regularised with a small inner core so that $\rho(0)$ is finite); and a cored Burkert profile~\cite{Burkert:1995yz}, $\rho\propto[(1+r/r_s)(1+(r/r_s)^2)]^{-1}$, which better fits dwarf-galaxy rotation curves. For more details on the cusp/core contrast we refer to~\cite{de2010core}.

\emph{Vainshtein-radius reduction} (Fig.~\ref{fig:robustness_halo}, top-right panel). For all three shapes the symbiotic Vainshtein radius lies below its pure-Galileon value throughout. The ratio equals the geometric value $\tilde\alpha^{-2/3}$ at low density and deepens as the growing core charge is increasingly screened (most for the flat-core profiles, least for the NFW-like cusp).

\emph{Chameleon-core de-screening} (Fig.~\ref{fig:robustness_halo}, bottom-right panel). The de-screening $\mathcal{D}$ [Eq.~\eqref{eq:Ddef}] grows with the Vainshtein strength $\beta$ for all profiles, confirming the effect is generic, though its onset depends on the core. The flat-core profiles (top-hat, cored) retain the thin-shell threshold (the core stays pinned, $\mathcal{D}=0$, until $\beta$ exceeds a critical value) whereas the NFW-like cusp, lacking a flat core to pin to, de-screens continuously. The idealised top-hat is thus the conservative case: realistic cuspy halos de-screen more readily.

\bibliographystyle{utphys}
\bibliography{Symbiosis_V+Ch}
\end{document}